\documentclass[preprint,prd,amsmath,amssymb]{revtex4}
\pdfoutput=1
\usepackage{graphicx}
\usepackage{color}
\newcommand{\beq}{\begin{eqnarray}}
\newcommand{\eeq}{\end{eqnarray}}

\newcommand{\bmp}{\noindent\begin{minipage}{16cm}}
\newcommand{\emp}{\end{minipage}\vskip 7mm} 

\usepackage{graphicx}
\usepackage{dcolumn}
\usepackage{bm}
\usepackage{amsmath}
\usepackage{amsfonts}
\usepackage{bbm}
\usepackage{subfigure}
\usepackage{pxfonts}
\usepackage{slashed}

\usepackage{ulem}

\newcommand{\mybar}[1]%
        {\kern 0.8pt\overline{\kern -0.8pt#1\kern -0.8pt}\kern 0.8pt}
\newcommand{\sla}[1]%
        {\raise.15ex\hbox{$/$}\kern-.57em #1}
\newcommand{\roughly}[1]%
        {\mathrel{\raise.3ex\hbox{$#1$\kern-.75em\lower1ex\hbox{$\sim$}}}}

\newcommand{\drawsquare}[2]{\hbox{%
\rule{#2pt}{#1pt}\hskip-#2pt
\rule{#1pt}{#2pt}\hskip-#1pt
\rule[#1pt]{#1pt}{#2pt}}\rule[#1pt]{#2pt}{#2pt}\hskip-#2pt
\rule{#2pt}{#1pt}}

\newcommand{\Yfund}{\raisebox{-.5pt}{\drawsquare{6.5}{0.4}}}
\newcommand{\Ysymm}{\raisebox{-.5pt}{\drawsquare{6.5}{0.4}}\hskip-0.4pt%
        \raisebox{-.5pt}{\drawsquare{6.5}{0.4}}}
\newcommand{\Yasymm}{\raisebox{-3.5pt}{\drawsquare{6.5}{0.4}}\hskip-6.9pt%
        \raisebox{3pt}{\drawsquare{6.5}{0.4}}}

\begin{document}
\title{\large  \color{red} Conformal Windows of SP(2N) and SO(N) Gauge Theories}
\author{Francesco {\sc Sannino}$^{\color{blue}{\varheartsuit}}$}
\email{sannino@ifk.sdu.dk} 
\affiliation{$^{\color{blue}{\varheartsuit}}$ 
University of Southern~Denmark,~Campusvej~55,~DK-5230~Odense M,~Denmark}


\begin{abstract}
We study the nonperturbative dynamics of nonsupersymmetric asymptotically free gauge theories with fermionic matter in distinct representations of  the SO(N) and Sp(2N) gauge groups. We use different analytic methods to unveil the associated conformal windows for the relevant matter representations. 
We propose a direct test for confronting and establishing the validity of the analytic methods used to constrain the conformal windows. By comparing the resulting windows for SU, Sp and SO a pleasing universal picture emerges. 
\end{abstract}

\maketitle

\section{Introduction}

Models of dynamical breaking of the electroweak symmetry are theoretically appealing and constitute one of the best motivated natural extensions of the Standard Model (SM).  These are also among the most challenging models to work with since they require deep knowledge of gauge dynamics in a regime where perturbation theory fails. In particular, it is of utmost importance to gain information on the nonperturbative dynamics of non-abelian four dimensional gauge theories. 

Recent studies of the dynamics of gauge theories featuring fermions transforming according to higher dimensional representations of the new gauge group led to several interesting phenomenological possibilities \cite{Sannino:2004qp,Dietrich:2005jn,Dietrich:2006cm} such as Minimal Walking Technicolor (MWT) \cite{Foadi:2007ue} and Ultra Minimal Walking Technicolor (UMT) \cite{Ryttov:2008xe}. Higher dimensional representations have been used earlier in particle physics phenomenology. Time honored examples are grand unified models. Theories with fermions transforming according to higher dimensional representations develop an infrared fixed point (IRFP) for a very small number of flavors and colors  \cite{Sannino:2004qp,Dietrich:2006cm,Ryttov:2007cx}. This was considered unlikely to occur for nonsupersymmetric gauge theories with fermionic matter \cite{Hill:2002ap}.  This discovery is important since it allows the construction of several explicit UV-complete models able to break the electroweak symmetry dynamically while naturally featuring small contributions to the electroweak precision parameters \cite{Appelquist:1998xf,Kurachi:2006mu,Foadi:2007ue}. Simultaneously it also helps alleviating the Flavor Changing Neutral Currents while the models also feature explicit candidates of asymmetric dark matter \cite{Foadi:2007ue,Ryttov:2008xe}. These models are economical since they require the introduction of a very small number of underlying elementary fields and can feature a light composite Higgs \cite{Dietrich:2005jn,Dietrich:2006cm,Hong:2004td}.  Recent analyses lend further support to the latter observation \cite{Doff:2009nk,Doff:2008xx}. 

At large distances theories developing an IRFP are conformal. One can envision several ways to depart from conformality. {} For example one can add a relevant operator such as an explicit fermion mass term  or decrease the number of flavors. If the departure from conformality is {\it soft}, meaning that the IRFP is quasi-reached the gauge coupling constant runs slowly over a long range of energies and the theory is said to   {\it walk}  
\cite{Eichten:1979ah,Holdom:1981rm,Yamawaki:1985zg,Appelquist:1986an}. This is, however, not the best way to define a walking theory since the coupling constant is not a physical quantity. In fact one should look at two and higher point correlators and determine the associated scaling exponent. In a (quasi)-conformal theory the scaling will have a characteristic power law behavior. 
Gauge theories developing an IRFP are natural ultraviolet completions of unparticle \cite{Georgi:2007ek} models \cite{Sannino:2008nv,Sannino:2008ha}. The effects of the instantons and their interplay with the fermion-mass operator on the conformal window have been evaluated in \cite{Sannino:2008pz}.  Within the SD approach these effects were investigated in \cite{Appelquist:1997dc}.

Non-abelian gauge theories exist in a number of  distinct phases which can be classified according to the characteristic dependence of the potential energy on the distance between 
two well separated static sources.  The collection of all of these different behaviors,
when represented, for example, in the  flavor-color space, constitutes the {\it phase diagram} of the given gauge theory. The phase diagram of $SU(N)$ gauge theories as functions of number of flavors, colors and matter representation has been investigated in \cite{Sannino:2004qp,Dietrich:2006cm,Ryttov:2007sr,Ryttov:2007cx,Sannino:2008ha}. Interesting applications have been envisioned not only for the LHC phenomenology \cite{Sannino:2004qp,Foadi:2007ue,Belyaev:2008yj,Christensen:2005cb,Gudnason:2006mk,Dietrich:2009ix} but also for Cosmology \cite{Nussinov:1985xr,Barr:1990ca,Foadi:2008qv,Ryttov:2008xe,Nardi:2008ix, Gudnason:2006yj,Kainulainen:2006wq,Kouvaris:2007iq,Cline:2008hr,Kouvaris:2008hc,Kikukawa:2007zk,Jarvinen:2009wr,Antipin:2009ch}. The nonperturbative dynamics of these models is being investigated via first principles lattice computations by several groups \cite{Catterall:2007yx,Catterall:2008qk,
Shamir:2008pb,DelDebbio:2008wb,DelDebbio:2008zf, Hietanen:2008vc,Hietanen:2008mr,Appelquist:2007hu,Deuzeman:2008sc,Fodor:2008hn}. In the literature the reader can also find various attempts to gain information on the nonperturbative gauge dynamics  using  gauge-gravity type duality and we cite here only a few recent efforts \cite{Hirn:2008tc,Dietrich:2008ni,Nunez:2008wi}.

Here we extend the analysis of  the zero temperature and matter density phase diagram to $SO(N)$ and $Sp(2N)$ gauge theories. Our results will lead to a deeper understanding of the (conformal) gauge dynamics of nonsupersymmetric gauge theories while it will enlarge the number of nonsupersymmetric gauge theories which can be used for extending the SM.

The analytical tools we will use for such an exploration are: i) The conjectured all-orders beta function for nonsupersymmetric gauge theories with fermionic matter in arbitrary representations of the gauge group \cite{Ryttov:2007cx}; ii) The truncated Schwinger-Dyson equation (SD) \cite{Appelquist:1988yc,Cohen:1988sq,Miransky:1996pd} (referred also as the ladder approximation in the literature);  The Appelquist-Cohen-Schmaltz (ACS) conjecture \cite{Appelquist:1999hr} which makes use of the counting of the thermal degrees of freedom at high and low temperature. 

We will show that relevant constraints can be deduced for any gauge theory and any representation only via the all-orders beta function and the SD results. The ACS conjecture is, unfortunately, not sufficiently constraining when studying theories with matter in higher dimensional representations of $SO$ and $Sp$ gauge theories. This is in complete agreement with our earlier results for $SU$ gauge theories \cite{Sannino:2005sk}. We will re-discuss the phase diagram of the $SU(2)$ gauge theory with fundamental fermions. The results, here, seem to disagree with the ones in \cite{Appelquist:1999hr}. We suggest that by investigating the dynamics of the $SU(2)$ gauge theory with five Dirac flavors in the fundamental representation of the underlying gauge theory via first principles lattice simulations one will be able to test the ACS conjecture as well as the all-orders beta function one. 

The paper is organized as follows. In section \ref{All-orders} we will introduce the all-orders beta function, in section \ref{ra} we will summarize the basic points about the Schwinger-Dyson (SD) approximation, and in section \ref{ACS} we will briefly summarize the thermal degrees of freedom method to bound the conformal window. The phase diagram of $Sp(2N)$ gauge theories with matter in the vector and two-index representation will be investigated in section \ref{sp} while in section \ref{so} we will investigate the one for $SO(N)$ gauge theories. We will conclude in section \ref{conclusion}.

\section{All-orders Beta Function - Conjecture}
\label{All-orders}
Recently we have conjectured an all-orders beta function which allows for a bound of the conformal window \cite{Ryttov:2007cx} of $SU(N)$ gauge theories for any matter representation.  It is  written in a form useful for constraining the phase diagram of strongly coupled theories. It is inspired by the Novikov-Shifman-Vainshtein-Zakharov  (NSVZ) beta function for supersymmetric theories \cite{Novikov:1983uc,Shifman:1986zi} and the renormalization scheme coincides with the NSVZ one. The predictions of the conformal window coming from the above beta function are nontrivially supported by all the recent lattice results \cite{Catterall:2007yx,DelDebbio:2008wb,Catterall:2008qk,Appelquist:2007hu,
Shamir:2008pb,Deuzeman:2008sc,Lucini:2007sa}.

It reproduces the exact supersymmetric results when reducing the matter content to the one of supersymmetric gauge theories. In particular we compared our prediction for the running of the coupling constant for the pure Yang-Mills theories with the one studied via the Schroedinger functional \cite{Luscher:1992zx,Luscher:1993gh,Lucini:2007sa} and found an impressive agreement. We have also predicted that the IRFP for $SU(3)$ gauge theories could not extend below $8.25$ number of flavors. Subsequent numerical analysis  \cite{Appelquist:2007hu, Deuzeman:2008sc,Appelquist:2009ty} confirmed our prediction. 

Here we further assume the form of the beta function to hold for $SO(N)$ and $Sp(2N)$ gauge groups. Consider a generic gauge group with $N_f(r_i)$ Dirac flavors belonging to the representation $r_i,\ i=1,\ldots,k$ of the gauge group. The conjectured beta function reads:
\begin{eqnarray}
\beta(g) &=&- \frac{g^3}{(4\pi)^2} \frac{\beta_0 - \frac{2}{3}\, \sum_{i=1}^k T(r_i)\,N_{f}(r_i) \,\gamma_i(g^2)}{1- \frac{g^2}{8\pi^2} C_2(G)\left( 1+ \frac{2\beta_0'}{\beta_0} \right)} \ ,
\end{eqnarray}
with
\begin{eqnarray}
\beta_0 =\frac{11}{3}C_2(G)- \frac{4}{3}\sum_{i=1}^k \,T(r_i)N_f(r_i) \qquad \text{and} \qquad \beta_0' = C_2(G) - \sum_{i=1}^k T(r_i)N_f(r_i)  \ .
\end{eqnarray}
The generators $T_r^a,\, a=1\ldots N^2-1$ of the gauge group in the
representation $r$ are normalized according to
$\text{Tr}\left[T_r^aT_r^b \right] = T(r) \delta^{ab}$ while the
quadratic Casimir $C_2(r)$ is given by $T_r^aT_r^a = C_2(r)I$. The
trace normalization factor $T(r)$ and the quadratic Casimir are
connected via $C_2(r) d(r) = T(r) d(G)$ where $d(r)$ is the
dimension of the representation $r$. The adjoint
representation is denoted by $G$.

The beta function is given in terms of the anomalous dimension of the fermion mass $\gamma=-{d\ln m}/{d\ln \mu}$ where $m$ is the renormalized mass, similar to the supersymmetric case \cite{Novikov:1983uc,Shifman:1986zi,Jones:1983ip}. 
The loss of asymptotic freedom is determined by the change of sign in the first coefficient $\beta_0$ of the beta function. This occurs when
\begin{eqnarray} \label{AF}
\sum_{i=1}^{k} \frac{4}{11} T(r_i) N_f(r_i) = C_2(G) \ , \qquad \qquad \text{Loss of AF.}
\end{eqnarray}
 At the zero of the beta function we have
\begin{eqnarray}
\sum_{i=1}^{k} \frac{2}{11}T(r_i)N_f(r_i)\left( 2+ \gamma_i \right) = C_2(G) \ ,
\end{eqnarray}
Hence, specifying the value of the anomalous dimensions at the IRFP yields the last constraint needed to construct the conformal window. Having reached the zero of the beta function the theory is conformal in the infrared. For a theory to be conformal the dimension of the non-trivial spinless operators must be larger than one in order not to contain negative norm states \cite{Mack:1975je,Flato:1983te,Dobrev:1985qv}.  Since the dimension of the chiral condensate is $3-\gamma_i$ we see that $\gamma_i = 2$, for all representations $r_i$, yields the maximum possible bound
\begin{eqnarray}\label{Bound}
\sum_{i=1}^{k} \frac{8}{11} T(r_i)N_f(r_i) = C_2(G) \ .
\end{eqnarray}
In the case of a single representation this constraint yields 
\begin{equation}
N_f(r)^{\rm BF} \geq \frac{11}{8} \frac{C_2(G)}{T(r)} \ .
\end{equation}
The actual size of the conformal window can be smaller than the one determined by the bound above, Eq. (\ref{AF}) and (\ref{Bound}). It may happen, in fact, that chiral symmetry breaking is triggered for a value of the anomalous dimension less than two. If this occurs the conformal window shrinks. Within the ladder approximation \cite{Appelquist:1988yc,{Cohen:1988sq}} one finds that chiral symmetry breaking occurs when the anomalous dimension is close to one. Picking $\gamma_i =1$ we find:
\begin{eqnarray}\label{One}
\sum_{i=1}^{k} \frac{6}{11} T(r_i)N_f(r_i) = C_2(G) \ .
\end{eqnarray}
When considering two distinct representations the conformal window becomes a three dimensional volume, i.e. the conformal volume \cite{Ryttov:2007sr}.  Of course, we recover the results by Banks and Zaks \cite{Banks:1981nn} valid in the perturbative regime of the conformal window.

\section{Schwinger-Dyson in the Rainbow Approximation}
\label{ra}
\begin{figure}[h]
\includegraphics{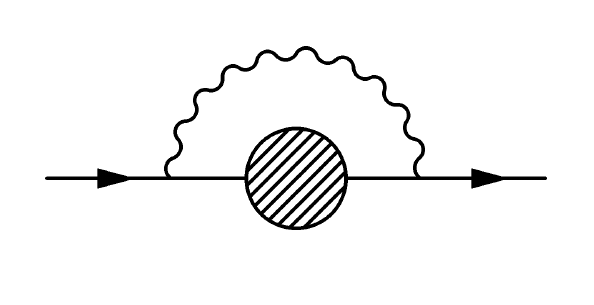}
\caption
{Rainbow approximation for the
fermion self energy function. The boson is a gluon.} 
\label{rainbowselfenergy}
\end{figure}

{}For nonsupersymmetric theories an old way to get quantitative estimates is to use the
{\it rainbow} approximation
to the Schwinger-Dyson equation
\cite{Pagels:1974se}, see Fig.~\ref{rainbowselfenergy}. 
\noindent
Here the full nonperturbative fermion propagator in momentum space reads
\beq iS^{-1}(p) = Z(p)\left(\slashed{p}-\Sigma(p)\right) \ , \eeq
and the Euclidianized gap equation in Landau gauge is given by
\beq \Sigma(p) =
3C_2(r)\int\frac{d^4k}{(2\pi)^4}\frac{\alpha\left((k-p)^2\right)}{(k-p)^2}\frac{\Sigma(k^2)}{Z(k^2)k^2
    + \Sigma^2(k^2)} \ , \eeq
where $Z(k^2)=1$ in the Landau gauge and we linearize
the equation by neglecting $\Sigma^2(k^2)$ in the denominator. Upon converting it into a differential equation and assuming that the coupling
$\alpha(\mu) \approx \alpha_c$ is varying slowly ($\beta(\alpha) \simeq
0$) one gets the   
approximate (WKB) 
solutions \cite{Fukuda:1976zb}
\beq \Sigma(p) \propto p^{-\gamma(\mu)} \ , \qquad \Sigma(p) \propto
p^{\gamma(\mu) - 2} \ , \label{sol-to-gap-eq} \eeq
where the critical coupling is given in terms of the quadratic Casimir
of the representation of the fermions
\beq \alpha_c \equiv \frac{\pi}{3C_2(r)} \ . \label{critical-coupling}
\eeq
The anomalous dimension of the fermion mass operator is
\beq \gamma(\mu) = 1 - \sqrt{1-\frac{\alpha(\mu)}{\alpha_c}} 
\sim \frac{3C_2(r)\alpha(\mu)}{2\pi} \ . \label{admp}
\eeq
The first solution
corresponds 
to the running of an ordinary mass term ({\it  hard} mass) of nondynamical origin and the
second solution  to a {\it soft} mass dynamically generated. In fact in the second case one observes the $1/p^2$ behavior in the limit of large momentum. 

Within this approximation spontaneous symmetry breaking occurs when $\alpha$ reaches the critical coupling $\alpha_c$ given in Eq.~(\ref{critical-coupling}). {}From Eq.~(\ref{admp}) it is clear that $ \alpha_c $ is reached when 
$\gamma$  is of order unity \cite{Cohen:1988sq,Appelquist:1988yc}. Hence the symmetry breaking occurs when the soft and the hard mass terms scale as function of the energy scale in the same way. In Ref.~\cite{Appelquist:1988yc},  it was noted that in the lowest (ladder) order, the gap equation leads to the condition $\gamma(2-\gamma)=1$ for chiral symmetry breaking to occur. To all orders in perturbation theory this condition is gauge invariant and also equivalent nonperturbatively to the condition $\gamma=1$. However, to any finite order in perturbation theory these conditions are, of course, different. Interestingly the condition $\gamma(2-\gamma)=1$ leads again to the critical coupling $\alpha_c$  when using the perturbative leading order expression for the anomalous dimension which is $\gamma=\frac{3C_2(r)}{2\pi}\alpha$ .

To summarize, the idea behind this method is simple. One simply compares the two couplings in the infrared associated to i) an infrared zero in the $\beta$ function, call it $\alpha^{\ast}$  with ii) the critical coupling, denoted with $\alpha_c$, above which a dynamical mass for the fermions generates nonperturbatively and chiral symmetry breaking occurs. If $\alpha^{\ast}$ is less than $\alpha_c$ chiral symmetry does not occur and the theory remains conformal in the infrared, viceversa if $\alpha^{\ast}$ is larger than $\alpha_c$ then the fermions acquire a dynamical mass and the theory cannot be conformal in the infrared. The condition $\alpha^{\ast} = \alpha_c$ provides the desired $N_f^{\rm SD} $ as function of $N$.  In practice to estimate $\alpha^{\ast}$ one uses the two-loop beta function while the truncated SD  equation to determine $\alpha_c$ as we have done before. This corresponds to when the anomalous dimension of the quark mass operator
becomes approximately unity.  

The two-loop fixed point value of the coupling constant
is:
\begin{eqnarray}
\frac{\alpha^*}{4\pi}=-\frac{\beta_0}{\beta_1}.
\end{eqnarray}
with the following definition of the two-loop beta function 
\begin{eqnarray}\beta (g) = -\frac{\beta_0}{(4\pi)^2} g^3 - \frac{\beta_1}{(4\pi)^4} g^5 \ ,
\label{perturbative}
\end{eqnarray}
where $g$ is the gauge coupling and the beta function coefficients are given by
\begin{eqnarray}
\beta_0 &=&\frac{11}{3}C_2(G)- \frac{4}{3}T(r)N_f \\
\beta_1 &=&\frac{34}{3} C_2^2(G)
- \frac{20}{3}C_2(G)T(r) N_f  - 4C_2(r) T(r) N_f  \ .\end{eqnarray}
To this order the two coefficients are universal,
i.e. do not depend on which renormalization group scheme one has used to determine them.
The perturbative expression for the anomalous dimension reads:
\begin{equation}
\gamma(g^2) = \frac{3}{2} C_2(r) \frac{g^2}{4\pi^2} + O(g^4) \ .
\end{equation}
With $\gamma =-{d\ln m}/{d\ln \mu}$ and $m$ the renormalized fermion mass.

For a fixed number of colors the critical number of flavors for which the
order of $\alpha^{\ast}$ and $\alpha_c$ changes is defined by imposing 
$\alpha^*{=}\alpha_c$, and it is given by
\begin{eqnarray}
{N_f^{\rm SD}} &=& \frac{17C_2(G)+66C_2(r)}{10C_2(G)+30C_2(r)}
\frac{C_2(G)}{T(r)} \ . \label{nonsusy}
\end{eqnarray}
Comparing with the previous result obtained using the all-orders beta function we see that it is the coefficient of $C_2(G)/T(r)$ which is different.

\section{Thermal Counting of the Degrees of Freedom - Conjecture}
\label{ACS}
The free energy can be seen as a device to count
the relevant degrees of freedom. It can be computed, exactly, in two regimes 
of a generic asymptotically free theory: the very hot and the very cold one. 

The zero-temperature theory of interest is
characterized using the quantity $f_{IR}$, related to the free energy by
\begin{equation}  \label{eq:firdef}
f_{IR} \equiv - \lim_{T\to 0} \frac{{\cal F}(T)}{T^4}
\frac{90}{\pi^2} \ ,
\end{equation}
where $T$ is the temperature and ${\cal F}$ is the conventionally
defined free energy per unit volume. The limit is well defined if
the theory has an IRFP. 
For the special case of an
infrared-free theory 
\begin{eqnarray}
f_{IR} = \sharp~~{\rm Real~Bosons}~+ ~\frac{7}{4}~\sharp~~{\rm Weyl-Fermions} \ .
\end{eqnarray}
The corresponding expression in the large $T$ limit is
\begin{equation}  \label{eq:fuvdef}
f_{UV} \equiv - \lim_{T\to \infty} \frac{{\cal F}(T)}{T^4}
\frac{90}{\pi^2}\ .
\end{equation}
This limit is well defined if the theory has an ultraviolet
fixed point. For an asymptotically free theory $f_{UV}$ counts the
underlying ultraviolet d.o.f. in a similar way.

In terms of these quantities, the conjectured inequality \cite{Appelquist:1999hr} for any
asymptotically free theory is
\begin{equation}  \label{eq:ineq}
f_{IR} \le f_{UV}\ .
\end{equation}
This inequality has not been proven but it was
shown to be consistent with known results and then used to derive new
constraints for several strongly coupled, vector-like gauge theories.
The ACS conjecture has been used also for chiral gauge theories \cite{Appelquist:1999vs}. There it 
was also found that to make definite predictions a stronger requirement is needed \cite{Appelquist:2000qg}. 
\section{Phase Diagram of $Sp(2N)$ Gauge Theories}
\label{sp}
$Sp(2N)$ is the subgroup of $SU(2N)$ which leaves the tensor
$J^{c_1 c_2} = ({\bf 1}_{N \times N} \otimes i \sigma_2)^{c_1 c_2}$
invariant. Irreducible tensors of $Sp(2N)$ must be traceless with respect to
$J^{c_1 c_2}$. 
Here we consider $Sp(2N)$ gauge theories with fermions transforming according to a given irreducible representation. Since $\pi^4\left[Sp(2N)\right] =Z_2$  there is a Witten topological anomaly \cite{Witten:1982fp} whenever the sum of the Dynkin indices of the various matter fields is odd. The adjoint of $Sp(2N)$ is the two-index symmetric tensor.

 \subsection{$Sp(2N)$ with Vector Fields}
 
Consider $2N_f$ Weyl fermions ${q_c^i}$ with ${c=1,\ldots,2N}$ and $i=1, \ldots , 2 N_f$  in the  fundamental representation of $Sp(2N)$. We have omitted the $SL(2,C)$ spinorial indices. We need an even number of flavors to avoid the Witten anomaly since the Dynkin index of the vector representation is equal to one. In the following Table we summarize the properties of the theory
\[ \begin{array}{|c|c|c|c|c|} \hline
{\rm Fields} &  \left[ Sp(2N) \right] & SU(2N_f) & T[r_i] & d[r_i] \\ \hline \hline
q &\Yfund & \Yfund & \frac{1}{2}& 2N   \\
G_{\mu}&{\rm Adj} = \Ysymm  &1&  N+1 & N(2N +1)  \\
 \hline \end{array} 
\]

\subsubsection{Chiral Symmetry Breaking}
The theory is asymptotically free for $N_f \leq 11(N+1)/2$ while the relevant gauge singlet mesonic degree of freedom is:
\begin{equation}
M^{[i,j]} = \epsilon^{\alpha \beta} q_{\alpha,c_1}^{[ i} q_{\beta,c_2}^{j]} J^{c_1c_2}\ . 
\end{equation}
If the number of flavors is smaller than the critical number of flavors above which the theory develops an IRFP we expect this operator to condense and to break $SU(2N_f)$ to the maximal diagonal subgroup which is $Sp(2N_f)$ leaving behind $2N_f^2  - N_f -1$ Goldstone bosons. Also, there exist no $Sp(2N)$ stable operators constructed using the invariant tensor $\epsilon^{c_1c_2,\ldots c_{2N}}$ since they will break up into mesons $M$. This is so since  the invariant tensor $\epsilon^{c_1c_2 \ldots c_{2N}}$ breaks up into sums of products of $J^{c_1c_2}$. 

\subsubsection{All-orders Beta Function}
A zero in the numerator of the all-orders beta function leads to the following value of the anomalous dimension of the mass operator at the IRFP: 
\begin{equation}
\gamma_{\Yfund} = \frac{11(N+1)}{N_f}  - 2 \ .
\end{equation}
Since the (mass) dimension of any scalar gauge singlet operator must be, by unitarity arguments, larger than one at the IRFP, this implies that $\gamma_{\Yfund} \leq 2$. Defining with $\gamma_{\Yfund}^{\ast}$ the maximal anomalous dimension above which the theory loses the IRFP the conformal window is:
\begin{equation}
 \frac{11}{4} (N+1) \leq  \frac{11}{2+\gamma_{\Yfund}^{\ast}} (N+1) \leq N_f \leq \frac{11}{2} (N+1) \ .
\end{equation}
{}For the first inequality we have taken the maximal value allowed for the anomalous dimension, i.e. $\gamma_{\Yfund}^{\ast} = 2$. 
\subsubsection{SD}

The estimate from the truncated SD analysis yields as critical value of Weyl flavors:
\begin{equation} 
N_f^{\rm SD} = \frac{2(1+N)(67+100N)}{35+50N}  \ .
\end{equation}
\subsubsection{Thermal Degrees of Freedoms}
In the UV we have $2N(2N+1)$ gauge bosons, where the extra factor of two comes from taking into account the two helicities of each massless gauge boson, and $4NN_f$ Weyl fermions. In the IR we have $2N_f^2  - N_f -1$ Goldstones and hence we have: 
\begin{equation}
f_{UV} = 2N(2N+1) + 7{NN_f} \ , \qquad f_{IR} = 2N_f^2  - N_f -1 \ .
\end{equation}
The number of flavors for which $f_{IR} = f_{UV}$ is 
\begin{equation}
N_f^{\rm Therm} = \frac{1+7N + \sqrt{3(3+10N+27N^2)}}{4} \ .
\end{equation}
No information can be obtained about the value of the anomalous dimension of the fermion bilinear at the fixed point. Assuming the conjecture to be valid the critical number of flavors  cannot exceed $N_f^{\rm Therm}$.

\begin{figure}[h]
\centerline
{\includegraphics[height=6cm,width=18cm]{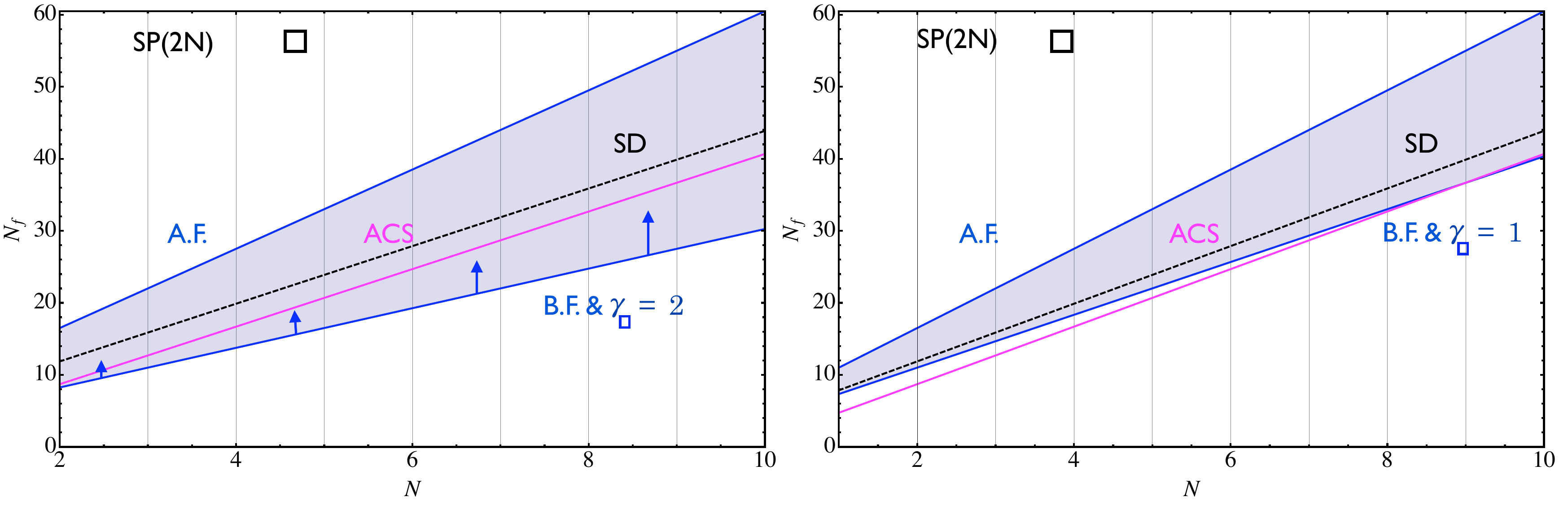}}
\caption
{Phase diagram of $Sp(2N)$ gauge theories with $2N_f$ fundamental Weyl fermions. {\it Left panel}: The upper solid (blue) line corresponds to the loss of asymptotic freedom and it is labeled by A.F.; the dashed (black) curve corresponds to the SD prediction for the breaking/restoring of chiral symmetry. The solid grey (magenta in color) line  corresponds to the ACS bound stating that the conformal region should start above this line. According to the all-orders beta function (B.F.) the conformal window cannot extend below the solid (blue) line, as indicated by the arrows. This line corresponds to the anomalous dimension of the mass 
reaching the maximum value of $2$.  {\it Right panel}: The B.F. line is plotted assuming  the value of the anomalous dimension to be one.} 
\label{Sp-Fundamental}
\end{figure}

\subsubsection{A comment on the limit  $N=1$ corresponding to $ SU(2)$}
In this case $N_f^{\rm Therm} = 2+\sqrt{\frac{15}{2}} \simeq 4.74$ and not $4\sqrt{4 - 16/81}\simeq 7.8$ as one deduces from equation (11) of \cite{Appelquist:1999hr}. The reason of the discrepancy is due to the fact that the fundamental representation of $SU(2) = Sp(2)$ is pseudoreal and hence the flavor symmetry is enhanced to $SU(2N_f)$. This enhanced symmetry is expected to break spontaneously to $Sp(2N_f)$. This yields $2N_f^2  - N_f -1$  Goldstone bosons rather than $N_f^2 - 1$ obtained assuming the global symmetry to be $SU(N_f)\times SU(N_f)\times U(1)$ spontaneously broken to $SU(N_f)\times U(1)$.  The corrected $N_f^{\rm Therm}$ value for $SU(2)$ is substantially lower than the $SD$ one which is $7.86$. The all-orders beta function result is instead $5.5$ for the lowest possible value of $N_f$ below which chiral symmetry must break (corresponding to $\gamma_{\Yfund} = 2$). Imposing $\gamma_{\Yfund} =1$ (suggested by the SD approach) the all-orders beta function returns $7.3$ which is closer to the $SD$ prediction. Note that there is some phenomenological interest in the $SU(2)$ gauge theory with fermionic matter in the fundamental representation. {}For example the case of $N_f=8$ has been employed in the literature as a possible template for early models of walking technicolor \cite{Appelquist:1997fp}. 
 
These results indicate that it is interesting to study the $SU(2)$ gauge theory with $N_f=5$ Dirac flavors via first principles Lattice simulation. This will allow to discriminate between the two distinct predictions, the one from the ACS and the one from the all-orders beta function.

\subsection{$Sp(2N)$ with Adjoint Matter Fields}

Consider $N_f$ Weyl fermions ${q_{\{ c_1,c_2 \}}^i}$ with $c_1$ and $c_2$ ranging from $1$ to $2N$ and $i=1, \ldots , N_f$. This is the  adjoint representation of $Sp(2N)$ with Dynkin index $2(N+1)$. Since it is even for any $N$ there is no Witten anomaly for any $N_f$. In the following Table we summarize the properties of the theory

 \[ \begin{array}{|c|c|c|c|c|} \hline
{\rm Fields} &  \left[ Sp(2N) \right] & SU(N_f) & T[r_i] & d[r_i] \\ \hline \hline
q &\Ysymm & \Yfund &  N+1&   N(2N +1)\\
G_{\mu}&{\rm Adj} = \Ysymm  &1&  N+1 & N(2N +1) \\
 \hline \end{array} 
\]

\subsubsection{Chiral Symmetry Breaking}
The theory is asymptotically free for $N_f \leq 11/2$ (recall that $N_f$ here is the number of Weyl fermions) while the relevant gauge singlet mesonic degree of freedom is:
\begin{equation}
M^{\{i,j \}} = \epsilon^{\alpha \beta} q_{\alpha, \{c_1,c_2\}}^{\{ i} q_{\beta, \{c_3,c_4 \}}^{j\}} J^{c_1c_3}J^{c_2c_4}\ . 
\end{equation}
If the number of flavors is smaller than the critical number of flavors above which the theory develops an IRFP we expect this operator to condense and to break $SU(N_f)$ to the maximal diagonal subgroup which is $SO(N_f)$ leaving behind $(N_f^2  + N_f -2)/2$ Goldstone bosons.

\subsubsection{All-orders Beta Function}

Here the anomalous dimension of the mass operator at the IRFP is: 
\begin{equation}
\gamma_{\Ysymm} = \frac{11}{N_f}  - 2 \ .
\end{equation}
Since the dimension of any scalar gauge singlet operator must be larger than one at the IRFP, this implies that $\gamma_{\Ysymm} \leq 2$. Defining with $\gamma_{\Ysymm}^{\ast}$ the maximal anomalous dimension above which the theory loses the IRFP the conformal window is:
\begin{equation}
 \frac{11}{4} \leq  \frac{11}{2+\gamma_{\Ysymm}^{\ast}}  \leq N_f \leq \frac{11}{2} \ .
\end{equation}
 
\subsubsection{SD}

The estimate from the truncated SD analysis yields as critical value of flavors:
\begin{equation} 
N_f^{\rm SD} =4.15  \ .
\end{equation}
\subsubsection{Thermal Degrees of Freedoms}
In the ultraviolet we have $2N(2N+1)$ gauge bosons and $N(2N+1)N_f$ Weyl fermions. In the IR we have $(N_f^2  + N_f -2)/2$  Goldstone bosons. Hence: 
\begin{equation}
f_{UV} = 2N(2N+1) + \frac{7}{4}{N(2N+1)N_f} \ , \qquad f_{IR} =\frac{ N_f^2  + N_f -2}{2} \ .
\end{equation}
The number of flavors for which $f_{IR} = f_{UV}$ is 
\begin{equation}
N_f^{\rm Therm} = \frac{-2 + 7 N + 14 N^2 + \sqrt{36 + 36 N + 121 N^2 + 196 N^3 + 196 N^4}}{4} \ .
\end{equation}
This is a monotonically increasing function of $N$ which even for a value of $N$ as low as 2 yields $N_f^{\rm Therm}=35.2$ which is several times higher than the limit set by asymptotic freedom. Although this fact does not contradict the statement that the critical number of flavors is lower than $N_f^{\rm Therm}$ it shows that this conjecture does not lead to useful constraints when looking at higher dimensional representations as we observed in \cite{Sannino:2005sk} when discussing higher dimensional representations for $SU(N)$ gauge groups.

\begin{figure}[h]
\centerline{
\includegraphics[height=6cm,width=18cm]{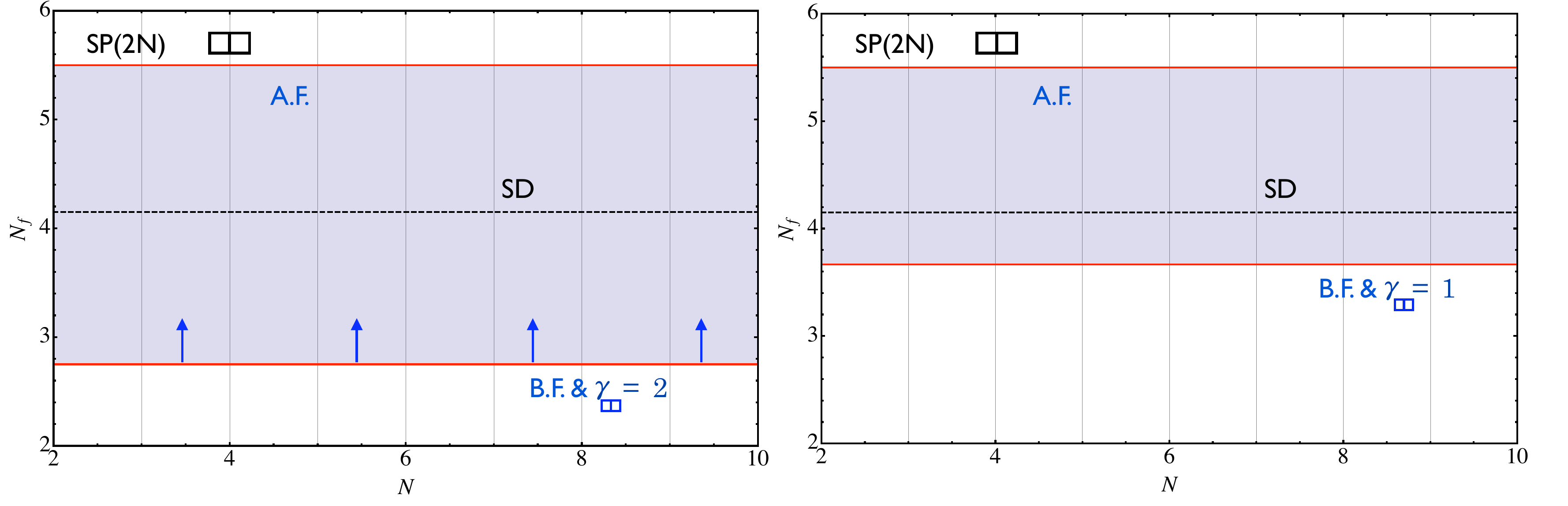}}
\caption
{Phase diagram of $Sp(2N)$ gauge theories with $N_f$ adjoint Weyl fermions. {\it Left panel}: The upper solid (red) line corresponds to the loss of asymptotic freedom and it is labeled by A.F.; the dashed (black) curve corresponds to the SD prediction for the breaking/restoring of chiral symmetry. According to the all-orders beta function (B.F.) the conformal window cannot extend below the solid (red) line, as indicated by the arrows. This line corresponds to the anomalous dimension of the mass 
reaching the maximum value of $2$.  {\it Right panel}: The B.F. line is plotted assuming  the value of the anomalous dimension to be one.} 
\label{Sp-Adj}
\end{figure}

 \subsection{$Sp(2N)$ with  Two-Index Anti-Symmetric Representation}

Consider $N_f$ Weyl fermions ${q_{[ c_1,c_2 ]}^i}$ with $c_1$ and $c_2$ ranging from $1$ to $2N$ and $i=1, \ldots , N_f$. As for the two-index symmetric case here too the Dynkin index is even  and hence we need not to worry about the Witten anomaly. In the following Table we summarize the properties of the theory
 \[ \begin{array}{|c|c|c|c|c|} \hline
{\rm Fields} &  \left[ Sp(2N) \right] & SU(N_f) & T[r_i] & d[r_i] \\ \hline \hline
q &\Yasymm & \Yfund &  N-1&N(2N-1) -1 \\
G_{\mu}&{\rm Adj} = \Ysymm  &1&  N+1 & N(2N +1)  \\
 \hline \end{array} 
\]

\subsubsection{Chiral Symmetry Breaking}
The theory is asymptotically free for $\displaystyle{N_f \leq \frac{11({N+1})}{2(N-1)}}$ with the relevant gauge singlet mesonic degree of freedom being:
\begin{equation}
M^{\{i,j \}} = \epsilon^{\alpha \beta} q_{\alpha, [c_1,c_2] }^{\{ i} q_{\beta, [c_3,c_4 ]}^{j\}} J^{c_1c_3}J^{c_2c_4}\ . 
\end{equation}
If the number of flavors is smaller than the critical number of flavors above which the theory develops an IRFP we expect this operator to condense and to break $SU(N_f)$ to the maximal diagonal subgroup which is $SO(N_f)$ leaving behind $(N_f^2  + N_f -2)/2$ Goldstone bosons.

\subsubsection{All-orders Beta Function}

The anomalous dimension of the mass operator at the IRFP is: 
\begin{equation}
\gamma_{\Yasymm} = \frac{11(N+1) -2N_f(N-1)} {N_f(N-1)}  .
\end{equation}
 Defining with $\gamma_{\Yasymm}^{\ast}$ the maximal anomalous dimension above which the theory loses the IRFP the conformal window is:
\begin{equation}
 \frac{11}{4} \frac{N+1}{N-1} \leq  \frac{11}{2+\gamma_{\Yasymm}^{\ast}} \frac{N+1}{N-1} \leq N_f \leq \frac{11}{2}   \frac{N+1}{N-1} \ .
\end{equation}
The maximal value allowed for the anomalous dimension is $\gamma_{\Yasymm}^{\ast} = 2$. 
\subsubsection{SD}

The SD analysis yields as critical value of flavors:
\begin{equation} 
N_f^{\rm SD} =\frac{(1+N) (83N+17)}{5(4N^2 - 3N - 1)}  \ .
\end{equation}

\subsubsection{Thermal Degrees of Freedoms}
In the ultraviolet we have $2N(2N+1)$ gauge bosons and $(N(2N-1)-1)N_f$ Weyl fermions. In the IR we have $(N_f^2  + N_f -2)/2$  Goldstone bosons. Hence: 
\begin{equation}
f_{UV} = 2N(2N+1) + \frac{7}{4}{(N(2N-1)-1)N_f} \ , \qquad f_{IR} =\frac{ N_f^2  + N_f -2}{2} \ .
\end{equation}
The number of flavors for which $f_{IR} = f_{UV}$ is 
\begin{equation}
N_f^{\rm Therm} = \frac{-9 - 7 N + 14 N^2 + \sqrt{113 + 190 N - 75 N^2 - 196 N^3 + 196 N^4}}{4} \ .
\end{equation}
As explained above no useful constraint can be set with this criterion \cite{Sannino:2005sk}.

\begin{figure}[h]
\centerline{
\includegraphics[height=6cm,width=18cm]{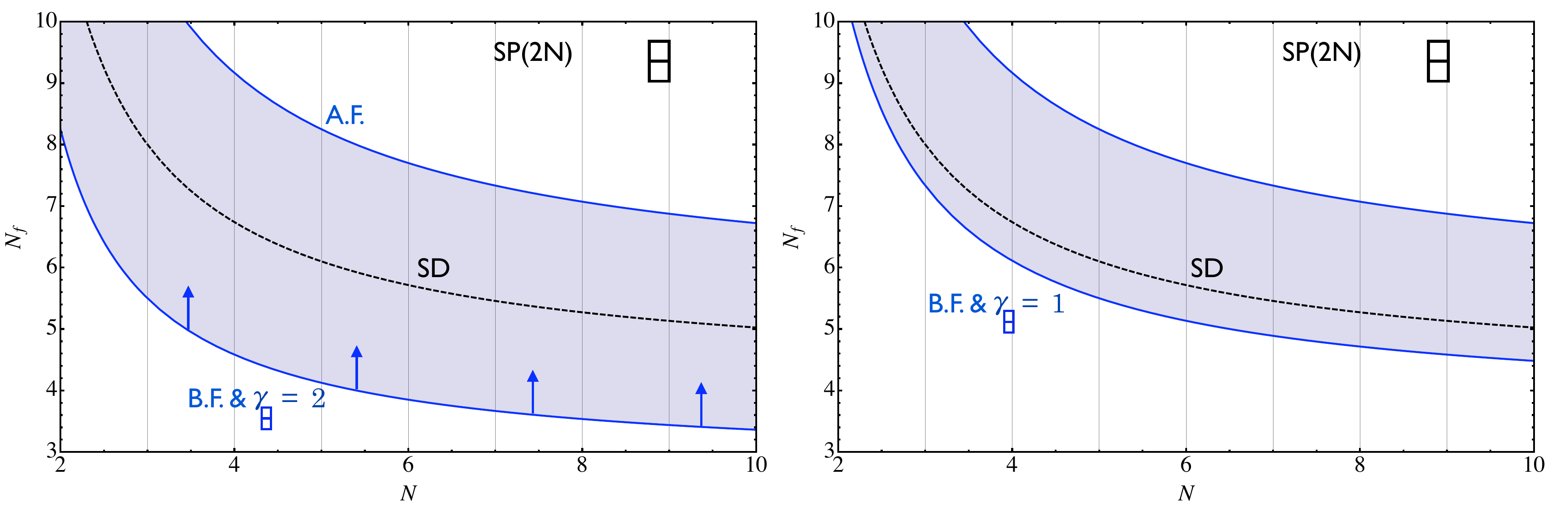}}
\caption
{Phase diagram of $Sp(2N)$ gauge theories with $N_f$ two-index antisymmetric Weyl fermions. {\it Left panel}: The upper solid (blue) curve corresponds to the loss of asymptotic freedom and it is labeled by A.F.; the dashed (black) curve corresponds to the SD prediction for the breaking/restoring of chiral symmetry. According to the all-orders beta function (B.F.) the conformal window cannot extend below the solid (blue) curve, as indicated by the arrows. This curve corresponds to the anomalous dimension of the mass 
reaching the maximum value of $2$.  {\it Right panel}: The B.F. curve is plotted assuming  the value of the anomalous dimension to be one.} 
\label{Sp-Ant}
\end{figure}

\subsubsection{Summary of the Results for $SP(2N)$ Gauge Theories}

\begin{figure}[h]
\centerline{
\includegraphics[height=8cm,width=13cm]{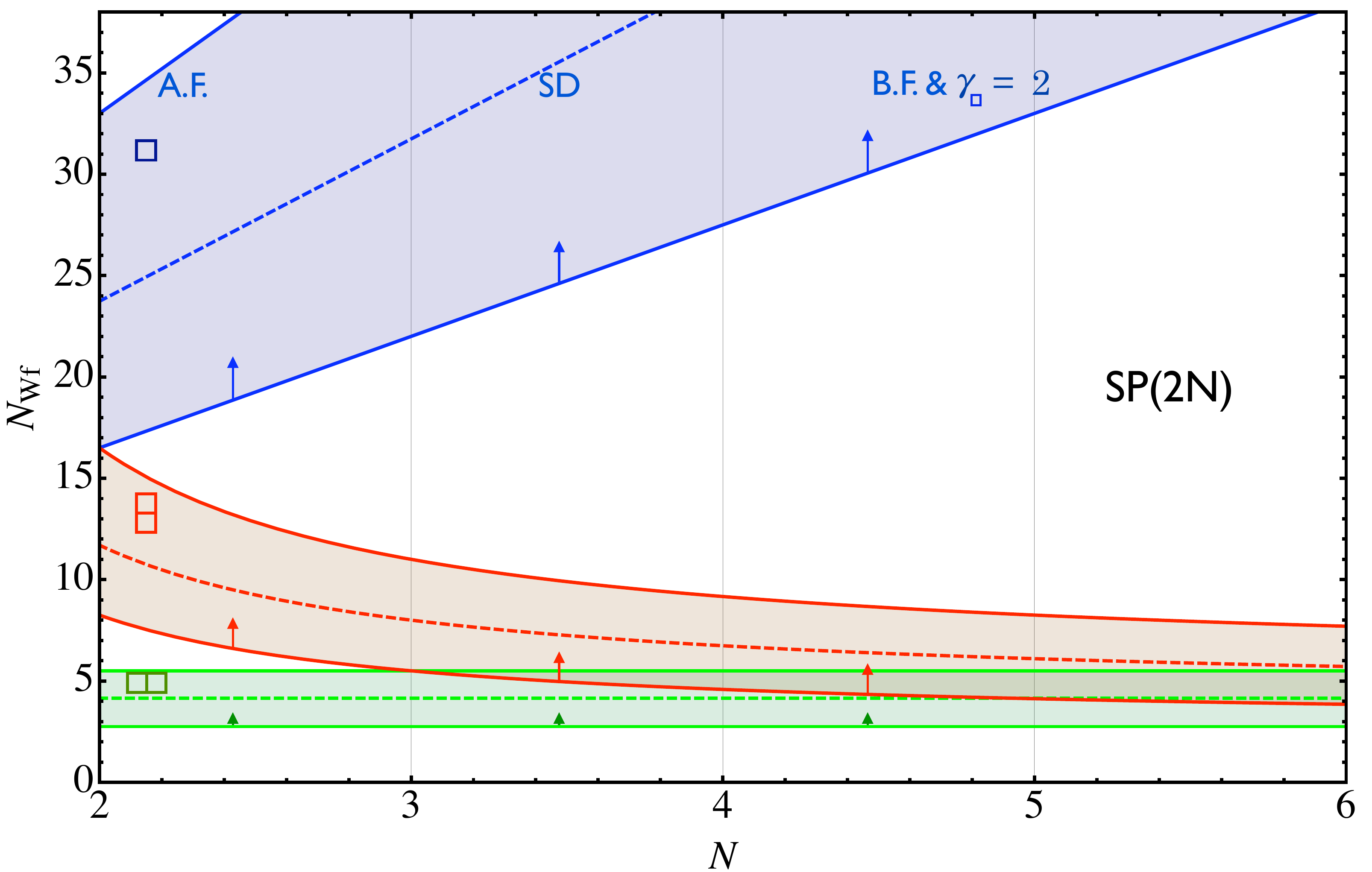}}
\caption
{Phase Diagram, from top to bottom, for $Sp(2N)$ Gauge Theories with $N_{Wf}=2N_f$  Weyl fermions in the vector representation (light blue),   $N_{Wf}=N_f$ in the  two-index antisymmetric representation (light red) and finally in the two-index symmetric (adjoint) (light green). The arrows indicate that the conformal windows can be smaller and the associated solid curves correspond to the all-orders beta function prediction for the maximum extension of the conformal windows.} 
\label{Sp-PhaseDiagram}
\end{figure}

In Figure~\ref{Sp-PhaseDiagram} we summarize the relevant zero temperature and matter density phase diagram as function of the number of colors and Weyl flavors ($N_{Wf}$) for $Sp(2N)$ gauge theories. For the vector representation $N_{Wf} = 2N_f$ while for the two-index theories $N_{Wf} = N_f$. The shape of the various conformal windows are very similar to the ones for $SU(N)$ gauge theories \cite{Sannino:2004qp,Dietrich:2006cm,Ryttov:2007cx} with the difference that in this case the two-index symmetric representation is the adjoint representation and hence there is one less conformal window.

\section{Phase Diagram of $SO(N)$ Gauge Theories}
\label{so}
We shall consider $SO(N)$ theories (for $N>5$) since they do not suffer of  a Witten
anomaly \cite{Witten:1982fp} and, besides, for $N<7$ can always be reduced to either an $SU$ or an $Sp$ theory.

 \subsection{$SO(N)$ with vector fields}
 
Consider $N_f$ Weyl fermions ${q_c^i}$ with ${c=1,\ldots,N}$ and $i=1, \ldots , N_f$  in the vector representation of $SO(N)$. In the following Table we summarize the properties of the theory

\[ \begin{array}{|c|c|c|c|c|} \hline
{\rm Fields} &  \left[ SO(N) \right] & SU(N_f) & T[r_i] & d[r_i] \\ \hline \hline
q &\Yfund & \Yfund & 1& N   \\
G_{\mu}&{\rm Adj} = \Yasymm  &1&  N-2 & \frac{N(N-1)}{2}  \\
 \hline \end{array} 
\]
\subsubsection{Chiral Symmetry Breaking}
The theory is asymptotically free for $\displaystyle{N_f \leq \frac{11({N-2})}{2}}$.  The relevant gauge singlet mesonic degree of freedom is:
\begin{equation}
M^{\{i,j \}} = \epsilon^{\alpha \beta} q_{\alpha, c_1 }^{\{ i} q_{\beta, c_2}^{j\}} \delta^{c_1c_2}\ . 
\end{equation}
If the number of flavors is smaller than the critical number of flavors above which the theory develops an IRFP we expect this operator to condense and to break $SU(N_f)$ to the maximal diagonal subgroup which is $SO(N_f)$ leaving behind $(N_f^2  + N_f -2)/2$ Goldstone bosons.

\subsubsection{All-orders Beta Function}

The anomalous dimension of the mass operator at the IRFP is: 
\begin{equation}
\gamma_{\Yfund} = \frac{11(N-2) } {N_f}  - 2 \  .
\end{equation}
Defining with $\gamma_{\Yfund}^{\ast}$ the maximal anomalous dimension above which the theory loses the IRFP the conformal window reads:
\begin{equation}
 \frac{11}{4} {N-2} \leq  \frac{11}{2+\gamma_{\Yfund}^{\ast}} {N-2} \leq N_f \leq \frac{11}{2}   {N-2} \ .
\end{equation}
The maximal value allowed for the anomalous dimension is $\gamma_{\Yfund}^{\ast} = 2$. 
\subsubsection{SD}

The SD analysis yields as critical value of flavors:
\begin{equation} 
N_f^{\rm SD} =\frac{2(N-2) (50N-67)}{5(5N - 7)}  \ .
\end{equation}

\subsubsection{Thermal Degrees of Freedoms}
In the ultraviolet we have $N(N- 1)$ gauge bosons and $NN_f$ Weyl fermions. In the IR we have $(N_f^2  + N_f -2)/2$  Goldstone bosons. Hence: 
\begin{equation}
f_{UV} = N(N-1) + \frac{7}{4}{NN_f} \ , \qquad f_{IR} =\frac{ N_f^2  + N_f -2}{2} \ .
\end{equation}
The number of flavors for which $f_{IR} = f_{UV}$ is 
\begin{equation}
N_f^{\rm Therm} = \frac{-2 + 7 N + \sqrt{36 - 60 N + 81 N^2}}{4} \ .
\end{equation}

This value is larger than the $SD$ result and it is larger than the asymptotic freedom constraint for $N<7$. This is not too surprising since the vector representation of $SO(N)$ for small $N$ becomes a higher representation of other groups for which we have already shown that this method is unconstraining  \cite{Sannino:2005sk}.

\begin{figure}[h]
\centerline{
\includegraphics[height=6cm,width=18cm]{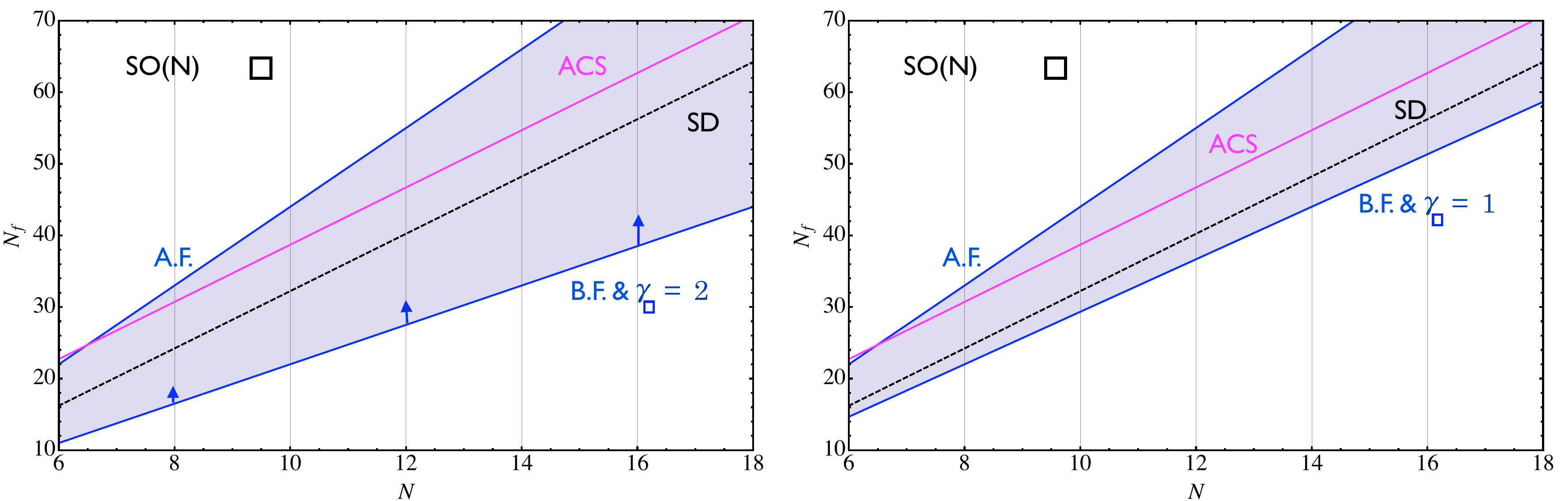}}
\caption
{Phase diagram of $SO(N)$ gauge theories with $N_f$ fundamental Weyl fermions. {\it Left panel}: The upper solid (blue) line corresponds to the loss of asymptotic freedom and it is labeled by A.F.; the dashed (black) curve corresponds to the SD prediction for the breaking/restoring of chiral symmetry. The solid grey (magenta in color) line  corresponds to the ACS bound stating that the conformal region should start above this line. According to the all-orders beta function (B.F.) the conformal window cannot extend below the solid (blue) line, as indicated by the arrows. This line corresponds to the anomalous dimension of the mass 
reaching the maximum value of $2$.  {\it Right panel}: The B.F. line is plotted assuming  the value of the anomalous dimension to be one.} 
\label{SO-Fundamental}
\end{figure}

Note that the ACS line is always above the SD result. 

 \subsection{$SO(N)$ with Adjoint Matter Fields}
  
Consider $N_f$ Weyl fermions ${q_{[c_1,c_2]}^i}$ with $c_1$ and $c_2$ varying in the range $1,\ldots,N $ and $i=1, \ldots , N_f$. This is the  adjoint representation of $SO(N)$. In the following Table we summarize the properties of the theory

 \[ \begin{array}{|c|c|c|c|c|} \hline
{\rm Fields} &  \left[ SO(N) \right] & SU(N_f) & T[r_i] & d[r_i] \\ \hline \hline
q &\Yasymm & \Yfund &  N-2& \frac{N(N-1)}{2}  \\
G_{\mu}&{\rm Adj} = \Yasymm  &1&  N-2 & \frac{N(N-1)}{2}  \\
 \hline \end{array} 
\]

The analysis leads to a conformal window which is an identical copy of the one for the adjoint matter of the $Sp$ gauge theory which is also identical to the $SU$ case with adjoint matter. 

 \subsection{$SO(N)$ with  Two-Index Symmetric Representation}

Consider $N_f$ Weyl fermions ${q_{\{c_1,c_2 \}}^i}$ with $c_1$ and $c_2$ varying in the range $1,\ldots,N $ and $i=1, \ldots , N_f$, i.e. in the two-index symmetric representation of $SO(N)$.  In the following Table we summarize the properties of the theory
 
 \[ \begin{array}{|c|c|c|c|c|} \hline
{\rm Fields} &  \left[ SO(N) \right] & SU(N_f) & T[r_i] & d[r_i] \\ \hline \hline
q &\Ysymm & \Yfund &  N+2& \frac{N(N+1)}{2} -1  \\
G_{\mu}&{\rm Adj} = \Yasymm  &1&  N-2 & \frac{N(N-1)}{2}  \\
 \hline \end{array} 
\]

\subsubsection{Chiral Symmetry Breaking}
The theory is asymptotically free for $\displaystyle{N_f \leq \frac{11({N-2})}{2(N+2)}}$.  The relevant gauge singlet mesonic degree of freedom is:
\begin{equation}
M^{\{i,j \}} = \epsilon^{\alpha \beta} q_{\alpha, \{c_1,c_2\} }^{\{ i} q_{\beta, \{c_3,c_4\}}^{j\}} \delta^{c_1c_3}\delta^{c_2,c_4}\ . 
\end{equation}
If the number of flavors is smaller than the critical number of flavors above which the theory develops an IRFP we expect this operator to condense and to break $SU(N_f)$ to the maximal diagonal subgroup which is $SO(N_f)$ leaving behind $(N_f^2  + N_f -2)/2$ Goldstone bosons.

\subsubsection{All-orders Beta Function}

The anomalous dimension of the mass operator at the IRFP is: 
\begin{equation}
\gamma_{\Ysymm} = \frac{11(N-2) } {N_f(N+2)}  - 2 \  .
\end{equation}
Defining with $\gamma_{\Ysymm}^{\ast}$ the maximal anomalous dimension above which the theory loses the IRFP the conformal window reads:
\begin{equation}
\frac{11}{4} \frac{N-2}{N+2} \leq  \frac{11}{2+\gamma_{\Ysymm}^{\ast}} \frac{N-2}{N+2} \leq N_f \leq \frac{11}{2}  \frac{N-2}{N+2} \ .
\end{equation}
The maximal value allowed for the anomalous dimension is $\gamma_{\Ysymm}^{\ast} = 2$. 
\subsubsection{SD}

The SD analysis yields as critical value of flavors:
\begin{equation} 
N_f^{\rm SD} =\frac{(N-2) (83N-34)}{10(2N^2 +3N -2)}  \ .
\end{equation}

\subsubsection{Thermal Degrees of Freedoms}
In the ultraviolet we have $N(N- 1)$ gauge bosons and $(N\frac{(N+1)}{2} -1)N_f$ Weyl fermions. In the IR we have $(N_f^2  + N_f -2)/2$  Goldstone bosons. Hence: 
\begin{equation}
f_{UV} = N(N-1) + \frac{7}{4}{(N\frac{(N+1)}{2} -1)N_f} \ , \qquad f_{IR} =\frac{ N_f^2  + N_f -2}{2} \ .
\end{equation}
The number of flavors for which $f_{IR} = f_{UV}$ is 
\begin{equation}
N_f^{\rm Therm} = \frac{-18 + 7 N(1+N) + \sqrt{452 + N (-380 + N (-75 + 49 N (2 + N)))}}{8} \ .
\end{equation}

This value is several times larger than the asymptotic freedom result and hence poses no constraint  \cite{Sannino:2005sk}.

\begin{figure}[h]
\centerline{
\includegraphics[height=6cm,width=18cm]{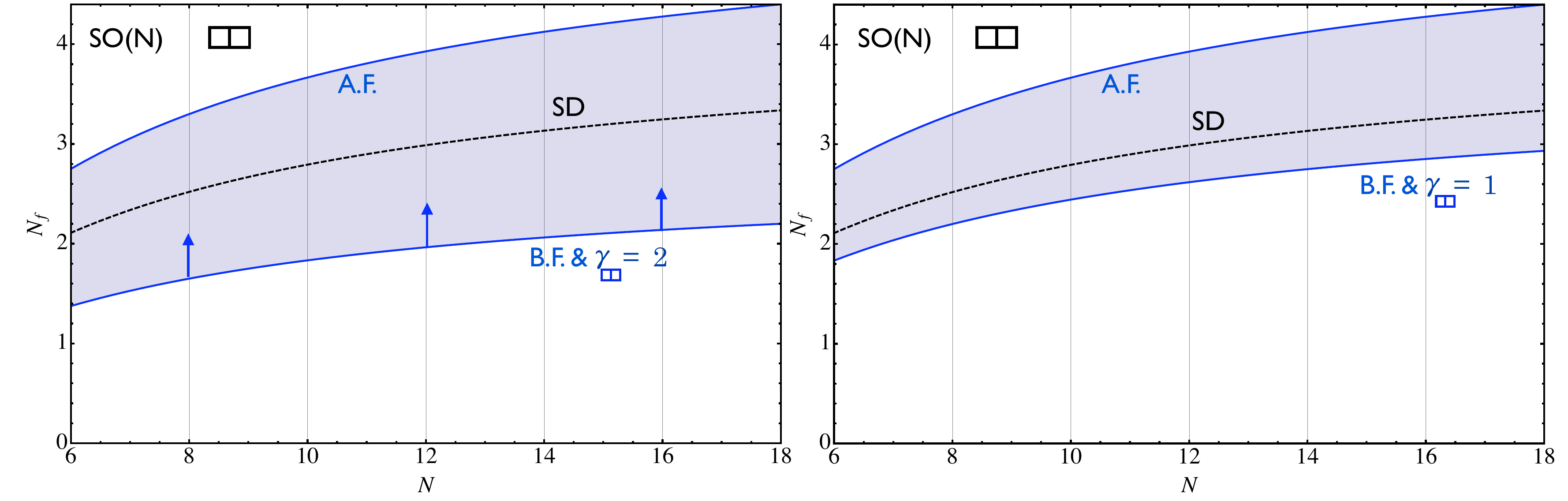}}
\caption
{Phase diagram of $SO(N)$ gauge theories with $N_f$ Weyl fermions in the two-index symmetric representation. {\it Left panel}: The upper solid (blue) curve corresponds to the loss of asymptotic freedom and it is labeled by A.F.; the dashed (black) curve corresponds to the SD prediction for the breaking/restoring of chiral symmetry. According to the all-orders beta function (B.F.) the conformal window cannot extend below the solid (blue) curve, as indicated by the arrows. This curve corresponds to the anomalous dimension of the mass 
reaching the maximum value of $2$.  {\it Right panel}: The B.F. curve is plotted assuming  the value of the anomalous dimension to be one.} 
\label{SO-Fundamental}
\end{figure}

\subsubsection{Summary for $SO(N)$ gauge theories}
\begin{figure}[t!]
\centerline{
\includegraphics[height=8cm,width=13cm]{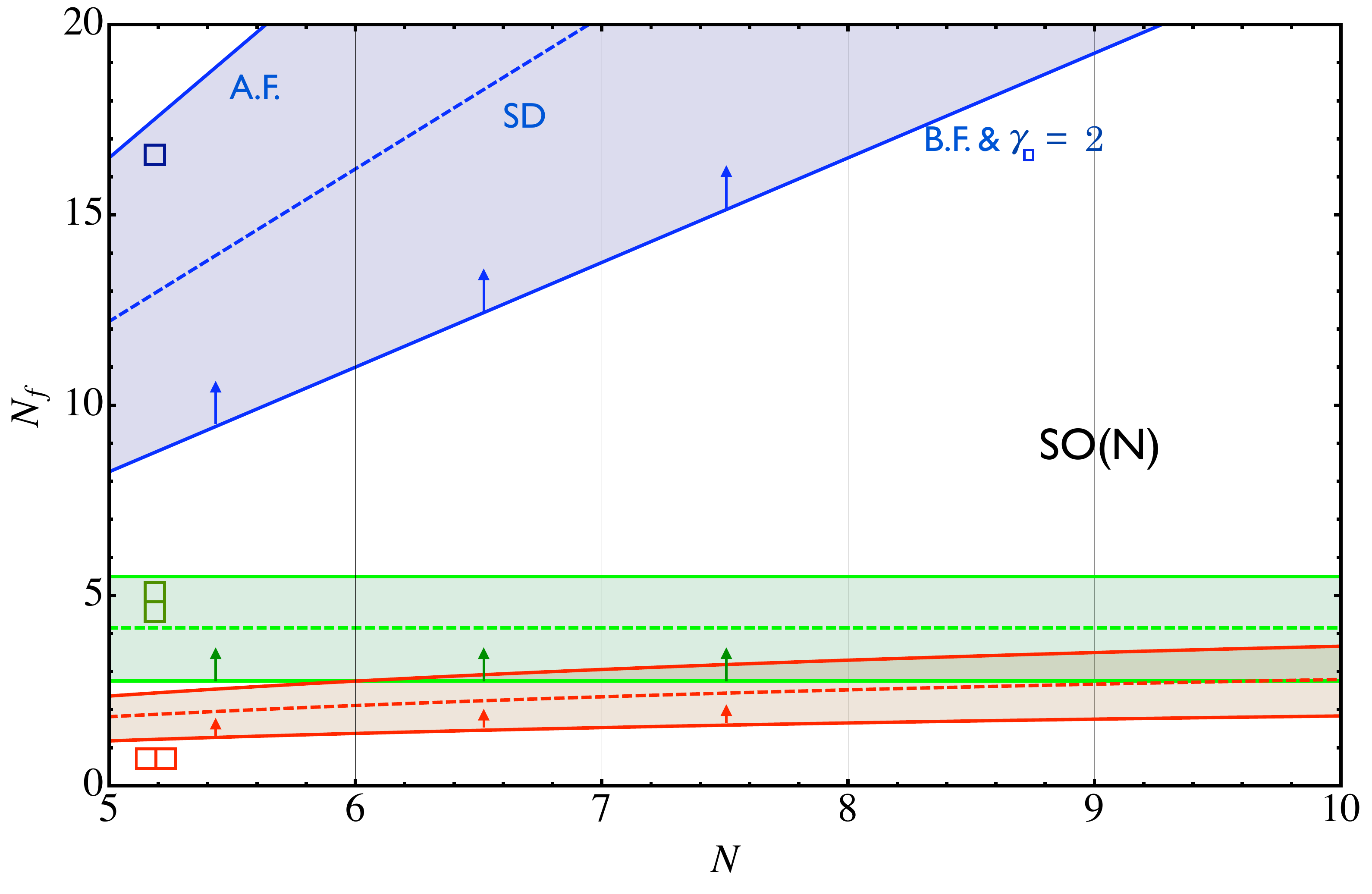}}
\caption
{Phase diagram of $SO(N)$ gauge theories with $N_f$  Weyl fermions in the vector representation, in the two-index antisymmetric (adjoint) and finally in the  two-index symmetric representation.  The arrows indicate that the conformal windows can be smaller and the associated solid curves correspond to the all-orders beta function prediction for the maximum extension of the conformal windows. } 
\label{So-PhaseDiagram}
\end{figure}

In the Fig.~\ref{So-PhaseDiagram} we summarize the relevant zero temperature and matter density phase diagram as function of the number of colors and Weyl flavors ($N_{f}$) for $SO(N)$ gauge theories. The shape of the various conformal windows are very similar to the ones for $SU(N)$  and $Sp(2N)$ gauge  with the difference that in this case the two-index antisymmetric representation is the adjoint representation. We have analyzed only the theories with $N\geq 6$ since the remaining smaller $N$ theories can be deduced from $Sp$ and $SU$ using the fact that 
$SO(6)\sim SU(4)$, $SO(5)\sim Sp(4)$, 
$SO(4)\sim SU(2)\times SU(2)$, $SO(3)\sim SU(2)$, and $SO(2)\sim U(1)$.  

 At infinite $N$ it is impossible to distinguish theories with matter in the two-index symmetric representation from theories with matter in the two-index antisymmetric. This means that, in this regime, one has an obvious equivalence between theories with these two types of matter. This statement is independent on whether the gauge group is $SU$, $Sp$ or $SO(N)$. What distinguishes $SU$ from both $Sp$ and $SO$ is the fact that in these two cases one of the two two-index representations is, in fact, the adjoint representation. This simple observation automatically implies that one Weyl flavor in the two-index symmetric (antisymmetric) representation of $SO(N)$($Sp(2N)$) becomes indistinguishable from pure super Yang-Mills at large N. The original observation appeared first within the context of string theory and it is due to Sugimoto \cite{Sugimoto:1999tx} and Uranga \cite{Uranga:1999ib}. A similar comment was made in \cite{Armoni:2007jt}.
 
\section{Comparison Chart and Conclusions}
\label{conclusion}
We unveiled the conformal windows for $SO$ and $Sp$ nonsupersymmetric gauge theories with fermions in the vector and two-index representations using three independent analytic methods. In Figures \ref{Sp-PhaseDiagram} and \ref{So-PhaseDiagram} we plotted the two phase diagrams as function of the number of flavors, colors and matter representation. These phase diagrams are similar to the one for $SU(N)$ gauge theories \cite{Sannino:2004qp,Dietrich:2006cm,Ryttov:2007cx} summarized in \cite{Sannino:2008ha}.  One observes a universal value, i.e. independent of the representation, of the ratio of the area of the maximum extension of the conformal window, predicted using the all-orders beta function, to the asymptotically free one, as defined in \cite{Ryttov:2007sr}. It is easy to check from our results that this ratio is not only independent on the representation but also on the particular gauge group chosen. 

The three different methods we used to unveil the conformal windows are the all-orders beta function (BF), the SD truncated equation and the thermal degrees of freedom method. In the Table below we compare directly the various analytical methods. 
\begin{table}[ht]
\caption{Direct comparison among the various analytic methods}
\centering
\begin{tabular}
{|c|c|c|c|c|c|} \hline
{Method} &  ~~~$\Yfund$ - Rep. ~~~& ~Higher Rep.~&~Multiple Rep.~&~~Susy~~& ~~~$\gamma$~~~  \\ \hline \hline
BF & +  & +  &+ & + & +  \\
SD &  + & +  & - & - & -\\
ACS & + & - & - & + &  - \\
 \hline 
 \end{tabular} 
 \label{comparison}
\end{table}
 The three plus signs in the second column indicate that the three analytic methods do constrain the conformal window of $SU$, $Sp$ and $SO$ gauge theories with fermions in the fundamental representation. Only BF and SD provide useful constraints in the case of the higher dimensional representations as summarized in the third column. {}When multiple representations participate in the gauge dynamics the BF constraints can be used directly \cite{Ryttov:2007cx,Ryttov:2008xe} to determine the extension of the conformal (hyper)volumes while extra dynamical information and approximations are required in the $SD$ approach. Since gauge theories with fermions in several representations of the underlying gauge group must contain higher dimensional representations the ACS is expected to be less efficient in this case \footnote{We do not consider super QCD a theory with higher dimensional representations.}. These results are summarized in the fourth column. The all-orders beta function reproduces the supersymmetric exact results when going over the super Yang-Mills case, the ACS conjecture was  proved successful when tested against the supersymmetric conformal window results \cite{Appelquist:1999hr}. However the SD approximation does not reproduce any supersymmetric result \cite{Appelquist:1997gq}.  The results are summarized in the fifth column. Finally, it is of  theoretical and phenomenological interest -- for example to construct sensible UV completions of models of dynamical electroweak symmetry breaking and unparticles --  to compute the anomalous dimension of the mass of the fermions at the (near) conformal fixed point.  Only the all-orders beta function provides a simple closed form expression as it is summarized in the sixth column. 

We have also suggested that it is interesting to study the $SU(2)$ gauge theory with $N_f=5$ Dirac flavors via first principles Lattice simulations since it will discriminate between the two distinct predictions, the one from the ACS conjecture and the one from the all-orders beta function.

Our analysis  substantially increases the number of asymptotically free gauge theories which can be used to construct SM extensions making use of (near) conformal dynamics. Current Lattice simulations can test our predictions and lend further support or even disprove  the emergence of a universal picture possibly relating the phase diagrams of gauge theories of fundamental interactions. 

\acknowledgments
 We thank T. Appelquist, S. Catterall,  D.D. Dietrich, R. Foadi, M.T. Frandsen, J. Giedt, M.O. J\"arvinen, T. Ryttov, H. Sakuma-Fukano, J. Schechter, R. Shrock and K.Tuominen for discussions and/or careful reading of the manuscript. The work is partially supported by the Marie Curie Excellence Grant under contract MEXT-CT-2004-013510 and the Major Framework Grant of the Danish Research Agency.



\begin{thebibliography}{99}

\bibitem{Sannino:2004qp}
  F.~Sannino and K.~Tuominen,
  `` Orientifold theory dynamics and symmetry breaking, ''
  Phys.\ Rev.\  D {\bf 71}, 051901 (2005)
  [arXiv:hep-ph/0405209].

\bibitem{Dietrich:2005jn}
  D.~D.~Dietrich, F.~Sannino and K.~Tuominen,
  ``Light composite Higgs from higher representations versus electroweak
  precision measurements: Predictions for LHC,''
  Phys.\ Rev.\  D {\bf 72}, 055001 (2005)
  [arXiv:hep-ph/0505059].

\bibitem{Dietrich:2006cm}
  D.~D.~Dietrich and F.~Sannino,
  `` Conformal window of SU(N) gauge theories with fermions in higher dimensional representations.''
  Phys.\ Rev.\  D {\bf 75}, 085018 (2007)
  [arXiv:hep-ph/0611341]. {}The section on the light composite Higgs is only on the archive version of the paper. 

\bibitem{Foadi:2007ue}
  R.~Foadi, M.~T.~Frandsen, T.~A.~Ryttov and F.~Sannino,
  ``Minimal Walking Technicolor: Set Up for Collider Physics,''
  Phys.\ Rev.\  D {\bf 76}, 055005 (2007)
  [arXiv:0706.1696 [hep-ph]].

\bibitem{Ryttov:2008xe}
  T.~A.~Ryttov and F.~Sannino,
  ``Ultra Minimal Technicolor and its Dark Matter TIMP,''
  Phys.\ Rev.\  D {\bf 78}, 115010 (2008)
  [arXiv:0809.0713 [hep-ph]].

\bibitem{Ryttov:2007cx}
  T.~A.~Ryttov and F.~Sannino,
  ``Supersymmetry Inspired QCD Beta Function,''
  Phys.\ Rev.\  D {\bf 78}, 065001 (2008)
  [arXiv:0711.3745 [hep-th]].

\bibitem{Hill:2002ap}
  C.~T.~Hill and E.~H.~Simmons,
  ``Strong dynamics and electroweak symmetry breaking,''
  Phys.\ Rept.\  {\bf 381}, 235 (2003)
  [Erratum-ibid.\  {\bf 390}, 553 (2004)]
  [arXiv:hep-ph/0203079].
  
\bibitem{Appelquist:1998xf}
  T.~Appelquist and F.~Sannino,
  ``The Physical Spectrum of Conformal SU(N) Gauge Theories,''
  Phys.\ Rev.\  D {\bf 59}, 067702 (1999)
  [arXiv:hep-ph/9806409].
  
\bibitem{Kurachi:2006mu}
  M.~Kurachi and R.~Shrock,
  ``Behavior of the S parameter in the crossover region between walking and
  QCD-like regimes of an SU(N) gauge theory,''
  Phys.\ Rev.\  D {\bf 74}, 056003 (2006)
  [arXiv:hep-ph/0607231].

\bibitem{Hong:2004td}
  D.~K.~Hong, S.~D.~H.~Hsu and F.~Sannino,
  ``Composite Higgs from higher representations,''
  Phys.\ Lett.\  B {\bf 597}, 89 (2004)
  [arXiv:hep-ph/0406200].

\bibitem{Doff:2009nk}
  A.~Doff and A.~A.~Natale,
  ``Mass and width of a composite Higgs boson,''
  arXiv:0902.2379 [hep-ph].

\bibitem{Doff:2008xx}
  A.~Doff, A.~A.~Natale and P.~S.~Rodrigues da Silva,
  ``Light composite Higgs from an effective action for technicolor,''
  Phys.\ Rev.\  D {\bf 77}, 075012 (2008)
  [arXiv:0802.1898 [hep-ph]].


\bibitem{Eichten:1979ah}
  E.~Eichten and K.~D.~Lane,
  ``Dynamical Breaking Of Weak Interaction Symmetries,''
  Phys.\ Lett.\  B {\bf 90}, 125 (1980).

\bibitem{Holdom:1981rm}
  B.~Holdom,
  ``Raising The Sideways Scale,''
  Phys.\ Rev.\  D {\bf 24}, 1441 (1981).

\bibitem{Yamawaki:1985zg}
  K.~Yamawaki, M.~Bando and K.~i.~Matumoto,
  ``Scale Invariant Technicolor Model And A Technidilaton,''
  Phys.\ Rev.\ Lett.\  {\bf 56}, 1335 (1986).

\bibitem{Appelquist:1986an}
  T.~W.~Appelquist, D.~Karabali and L.~C.~R.~Wijewardhana,
  ``Chiral Hierarchies and the Flavor Changing Neutral Current Problem in
  Technicolor,''
  Phys.\ Rev.\ Lett.\  {\bf 57}, 957 (1986).

\bibitem{Georgi:2007ek}
  H.~Georgi,
  ``Unparticle Physics,''
  Phys.\ Rev.\ Lett.\  {\bf 98}, 221601 (2007)
  [arXiv:hep-ph/0703260].

\bibitem{Sannino:2008nv}
  F.~Sannino and R.~Zwicky,
  ``Unparticle \& Higgs as Composites,''
  arXiv:0810.2686 [hep-ph].

\bibitem{Sannino:2008ha}
  F.~Sannino,
  ``Dynamical Stabilization of the Fermi Scale: Phase Diagram of Strongly
  Coupled Theories for (Minimal) Walking Technicolor and Unparticles,''
  arXiv:0804.0182 [hep-ph].

\bibitem{Sannino:2008pz}
  F.~Sannino,
  ``Conformal Chiral Dynamics,''
  arXiv:0811.0616 [hep-ph].

\bibitem{Appelquist:1997dc}
  T.~Appelquist and S.~B.~Selipsky,
  ``Instantons and the chiral phase transition,''
  Phys.\ Lett.\  B {\bf 400}, 364 (1997)
  [arXiv:hep-ph/9702404].

\bibitem{Ryttov:2007sr}
  T.~A.~Ryttov and F.~Sannino,
  ``Conformal Windows of SU(N) Gauge Theories, Higher Dimensional
  Representations and The Size of The Unparticle World,''
  Phys.\ Rev.\  D {\bf 76}, 105004 (2007)
  [arXiv:0707.3166 [hep-th]].



\bibitem{Belyaev:2008yj}
  A.~Belyaev, R.~Foadi, M.~T.~Frandsen, M.~Jarvinen, F.~Sannino and A.~Pukhov,
  ``Technicolor Walks at the LHC,''
  arXiv:0809.0793 [hep-ph].

\bibitem{Christensen:2005cb}
  N.~D.~Christensen and R.~Shrock,
  ``Technifermion representations and precision electroweak constraints,''
  Phys.\ Lett.\  B {\bf 632}, 92 (2006)
  [arXiv:hep-ph/0509109].

\bibitem{Gudnason:2006mk}
  S.~B.~Gudnason, T.~A.~Ryttov and F.~Sannino,
  ``Gauge coupling unification via a novel technicolor model,''
  Phys.\ Rev.\  D {\bf 76}, 015005 (2007)
  [arXiv:hep-ph/0612230].

\bibitem{Dietrich:2009ix}
  D.~D.~Dietrich and M.~Jarvinen,
  ``Pion masses in quasiconformal gauge field theories,''
  arXiv:0901.3528 [hep-ph].

\bibitem{Nussinov:1985xr}
  S.~Nussinov,
  ``Technocosmology: Could A Technibaryon Excess Provide A 'Natural' Missing
  Mass Candidate?,''
  Phys.\ Lett.\  B {\bf 165}, 55 (1985).

\bibitem{Barr:1990ca}
  S.~M.~Barr, R.~S.~Chivukula and E.~Farhi,
``Electroweak fermion number violation and the production of stable particles in the early universe,''
  Phys.\ Lett.\  B {\bf 241}, 387 (1990).

\bibitem{Foadi:2008qv}
  R.~Foadi, M.~T.~Frandsen and F.~Sannino,
  ``Technicolor Dark Matter,''
  arXiv:0812.3406 [hep-ph].

\bibitem{Nardi:2008ix}
  E.~Nardi, F.~Sannino and A.~Strumia,
  ``Decaying Dark Matter can explain the electron/positron excesses,''
  JCAP {\bf 0901}, 043 (2009)
  [arXiv:0811.4153 [hep-ph]].

\bibitem{Gudnason:2006yj}
  S.~B.~Gudnason, C.~Kouvaris and F.~Sannino,
  ``Dark Matter from new Technicolor Theories,''
  Phys.\ Rev.\  D {\bf 74}, 095008 (2006)
  [arXiv:hep-ph/0608055].

\bibitem{Kainulainen:2006wq}
  K.~Kainulainen, K.~Tuominen and J.~Virkajarvi,
  ``The WIMP of a minimal technicolor theory,''
  Phys.\ Rev.\  D {\bf 75}, 085003 (2007)
  [arXiv:hep-ph/0612247].

\bibitem{Kouvaris:2007iq}
  C.~Kouvaris,
  ``Dark Majorana Particles from the Minimal Walking Technicolor,''
  Phys.\ Rev.\  D {\bf 76}, 015011 (2007)
  [arXiv:hep-ph/0703266].
  
\bibitem{Cline:2008hr}
  J.~M.~Cline, M.~Jarvinen and F.~Sannino,
  ``The Electroweak Phase Transition in Nearly Conformal Technicolor,''
  Phys.\ Rev.\  D {\bf 78}, 075027 (2008)
  [arXiv:0808.1512 [hep-ph]].

\bibitem{Kikukawa:2007zk}
  Y.~Kikukawa, M.~Kohda and J.~Yasuda,
  ``First-order restoration of SU(Nf) x SU(Nf) chiral symmetry with large Nf
  and Electroweak phase transition,''
  Phys.\ Rev.\  D {\bf 77}, 015014 (2008)
  [arXiv:0709.2221 [hep-ph]].

\bibitem{Kouvaris:2008hc}
  C.~Kouvaris,
  ``The Dark Side of Strong Coupled Theories,''
  Phys.\ Rev.\  D {\bf 78}, 075024 (2008)
  [arXiv:0807.3124 [hep-ph]].
  
\bibitem{Jarvinen:2009wr}
  M.~Jarvinen, T.~A.~Ryttov and F.~Sannino,
  ``Extra Electroweak Phase Transitions from Strong Dynamics,''
  arXiv:0901.0496 [hep-ph].

\bibitem{Antipin:2009ch}
  O.~Antipin and K.~Tuominen,
  ``Discriminating between technicolor and warped extra dimensional model via
  pp $\to$ ZZ channel,''
  arXiv:0901.4243 [hep-ph].

\bibitem{Catterall:2007yx}
  S.~Catterall and F.~Sannino,
  ``Minimal walking on the lattice,''
  Phys.\ Rev.\  D {\bf 76}, 034504 (2007)
  [arXiv:0705.1664 [hep-lat]].

\bibitem{Catterall:2008qk}
  S.~Catterall, J.~Giedt, F.~Sannino and J.~Schneible,
  ``Phase diagram of SU(2) with 2 flavors of dynamical adjoint quarks,''
  JHEP {\bf 0811}, 009 (2008)
  [arXiv:0807.0792 [hep-lat]].

\bibitem{Shamir:2008pb}
  Y.~Shamir, B.~Svetitsky and T.~DeGrand,
  ``Zero of the discrete beta function in SU(3) lattice gauge theory with color
  sextet fermions,''
  Phys.\ Rev.\  D {\bf 78}, 031502 (2008)
  [arXiv:0803.1707 [hep-lat]].

\bibitem{DelDebbio:2008wb}
  L.~Del Debbio, M.~T.~Frandsen, H.~Panagopoulos and F.~Sannino,
  ``Higher representations on the lattice: perturbative studies,''
  JHEP {\bf 0806}, 007 (2008)
  [arXiv:0802.0891 [hep-lat]].

\bibitem{DelDebbio:2008zf}
  L.~Del Debbio, A.~Patella and C.~Pica,
  ``Higher representations on the lattice: numerical simulations. SU(2) with
  adjoint fermions,''
  arXiv:0805.2058 [hep-lat].

\bibitem{Hietanen:2008vc}
  A.~Hietanen, J.~Rantaharju, K.~Rummukainen and K.~Tuominen,
  ``Spectrum of SU(2) gauge theory with two fermions in the adjoint
  representation,''
  PoS {\bf LATTICE2008}, 065 (2008)
  [arXiv:0810.3722 [hep-lat]].

\bibitem{Hietanen:2008mr}
  A.~J.~Hietanen, J.~Rantaharju, K.~Rummukainen and K.~Tuominen,
 ``Spectrum of SU(2) lattice gauge theory with two adjoint Dirac flavours,''
  arXiv:0812.1467 [hep-lat].

\bibitem{Appelquist:2007hu}
  T.~Appelquist, G.~T.~Fleming and E.~T.~Neil,
  ``Lattice Study of the Conformal Window in QCD-like Theories,''
  Phys.\ Rev.\ Lett.\  {\bf 100}, 171607 (2008)
  [arXiv:0712.0609 [hep-ph]].

\bibitem{Deuzeman:2008sc}
  A.~Deuzeman, M.~P.~Lombardo and E.~Pallante,
  ``The physics of eight flavours,''
  Phys.\ Lett.\  B {\bf 670}, 41 (2008)
  [arXiv:0804.2905 [hep-lat]].

\bibitem{Fodor:2008hn}
  Z.~Fodor, K.~Holland, J.~Kuti, D.~Nogradi and C.~Schroeder,
  ``Probing technicolor theories with staggered fermions,''
  arXiv:0809.4890 [hep-lat].


\bibitem{Hirn:2008tc}
  J.~Hirn, A.~Martin and V.~Sanz,
  ``Describing viable technicolor scenarios,''
  Phys.\ Rev.\  D {\bf 78}, 075026 (2008)
  [arXiv:0807.2465 [hep-ph]].

\bibitem{Dietrich:2008ni}
  D.~D.~Dietrich and C.~Kouvaris,
  ``Constraining vectors and axial-vectors in walking technicolour by a
  holographic principle,''
  Phys.\ Rev.\  D {\bf 78}, 055005 (2008)
  [arXiv:0805.1503 [hep-ph]].


\bibitem{Nunez:2008wi}
  C.~Nunez, I.~Papadimitriou and M.~Piai,
  ``Walking Dynamics from String Duals,''
  arXiv:0812.3655 [hep-th].

\bibitem{Appelquist:2009ty}
  T.~Appelquist, G.~T.~Fleming and E.~T.~Neil,
  ``Lattice Study of Conformal Behavior in SU(3) Yang-Mills Theories,''
  arXiv:0901.3766 [hep-ph].

\bibitem{Appelquist:1988yc}
  T.~Appelquist, K.~D.~Lane and U.~Mahanta,
  ``On the ladder approximation for spontaneous chiral symmetry breaking,''
  Phys.\ Rev.\ Lett.\  {\bf 61}, 1553 (1988).

\bibitem{Cohen:1988sq}
  A.~G.~Cohen and H.~Georgi,
  ``Walking Beyond The Rainbow,''
  Nucl.\ Phys.\  B {\bf 314}, 7 (1989).

\bibitem{Miransky:1996pd}
  V.~A.~Miransky and K.~Yamawaki,
  ``Conformal phase transition in gauge theories,''
  Phys.\ Rev.\  D {\bf 55}, 5051 (1997)
  [Erratum-ibid.\  D {\bf 56}, 3768 (1997)]
  [arXiv:hep-th/9611142].

\bibitem{Appelquist:1999hr}
  T.~Appelquist, A.~G.~Cohen and M.~Schmaltz,
  ``A new constraint on strongly coupled field theories,''
  Phys.\ Rev.\  D {\bf 60}, 045003 (1999)
  [arXiv:hep-th/9901109].

\bibitem{Sannino:2005sk}
  F.~Sannino,
  ``Higher representations: Confinement and large N,''
  Phys.\ Rev.\  D {\bf 72}, 125006 (2005)
  [arXiv:hep-th/0507251].

\bibitem{Novikov:1983uc}
  V.~A.~Novikov, M.~A.~Shifman, A.~I.~Vainshtein and V.~I.~Zakharov,
  ``Exact Gell-Mann-Low Function Of Supersymmetric Yang-Mills Theories From
  Instanton Calculus,''
  Nucl.\ Phys.\  B {\bf 229}, 381 (1983).

\bibitem{Shifman:1986zi}
  M.~A.~Shifman and A.~I.~Vainshtein,
  ``Solution of the Anomaly Puzzle in SUSY Gauge Theories and the Wilson
  Operator Expansion,''
  Nucl.\ Phys.\  B {\bf 277}, 456 (1986)
  [Sov.\ Phys.\ JETP {\bf 64}, 428 (1986\ ZETFA,91,723-744.1986)].

\bibitem{Lucini:2007sa}
  B.~Lucini and G.~Moraitis,
  ``Determination of the running coupling in pure SU(4) Yang-Mills theory,''
  PoS {\bf LAT2007}, 058 (2007)
  [arXiv:0710.1533 [hep-lat]].

\bibitem{Luscher:1992zx}
  M.~Luscher, R.~Sommer, U.~Wolff and P.~Weisz,
  ``Computation Of The Running Coupling In The SU(2) Yang-Mills Theory,''
  Nucl.\ Phys.\  B {\bf 389}, 247 (1993)
  [arXiv:hep-lat/9207010].

\bibitem{Luscher:1993gh}
  M.~Luscher, R.~Sommer, P.~Weisz and U.~Wolff,
  ``A Precise determination of the running coupling in the SU(3) Yang-Mills
  theory,''
  Nucl.\ Phys.\  B {\bf 413}, 481 (1994)
  [arXiv:hep-lat/9309005].

\bibitem{Appelquist:2009ty}
  T.~Appelquist, G.~T.~Fleming and E.~T.~Neil,
  ``Lattice Study of Conformal Behavior in SU(3) Yang-Mills Theories,''
  arXiv:0901.3766 [hep-ph].

\bibitem{Jones:1983ip}
  D.~R.~T.~Jones,
  ``More On The Axial Anomaly In Supersymmetric Yang-Mills Theory,''
  Phys.\ Lett.\  B {\bf 123}, 45 (1983).

\bibitem{Mack:1975je}
  G.~Mack,
  ``All Unitary Ray Representations Of The Conformal Group SU(2,2) With
  Positive Energy,''
  Commun.\ Math.\ Phys.\  {\bf 55}, 1 (1977).

\bibitem{Flato:1983te}
  M.~Flato and C.~Fronsdal,
  ``Representations Of Conformal Supersymmetry,''
  Lett.\ Math.\ Phys.\  {\bf 8}, 159 (1984).

\bibitem{Dobrev:1985qv}
  V.~K.~Dobrev and V.~B.~Petkova,
  ``All Positive Energy Unitary Irreducible Representations Of Extended
  Conformal Supersymmetry,''
  Phys.\ Lett.\  B {\bf 162}, 127 (1985).

\bibitem{Banks:1981nn}
  T.~Banks and A.~Zaks,
  ``On The Phase Structure Of Vector-Like Gauge Theories With Massless
  Fermions,''
  Nucl.\ Phys.\  B {\bf 196}, 189 (1982).

\bibitem{Pagels:1974se}
  H.~Pagels,
  ``Departures From Chiral Symmetry: A Review,''
  Phys.\ Rept.\  {\bf 16}, 219 (1975).

\bibitem{Fukuda:1976zb}
  R.~Fukuda and T.~Kugo,
  ``Schwinger-Dyson Equation For Massless Vector Theory And Absence Of Fermion
  Pole,''
  Nucl.\ Phys.\  B {\bf 117}, 250 (1976).


\bibitem{Appelquist:1999vs}
  T.~Appelquist, A.~G.~Cohen, M.~Schmaltz and R.~Shrock,
  ``New constraints on chiral gauge theories,''
  Phys.\ Lett.\  B {\bf 459}, 235 (1999)
  [arXiv:hep-th/9904172].

\bibitem{Appelquist:2000qg}
  T.~Appelquist, Z.~y.~Duan and F.~Sannino,
  ``Phases of chiral gauge theories,''
  Phys.\ Rev.\  D {\bf 61}, 125009 (2000)
  [arXiv:hep-ph/0001043].

\bibitem{Witten:1982fp}
  E.~Witten,
  ``An SU(2) anomaly,''
  Phys.\ Lett.\  B {\bf 117}, 324 (1982).

\bibitem{Appelquist:1997fp}
  T.~Appelquist, J.~Terning and L.~C.~R.~Wijewardhana,
  ``Postmodern technicolor,''
  Phys.\ Rev.\ Lett.\  {\bf 79}, 2767 (1997)
  [arXiv:hep-ph/9706238].


\bibitem{Appelquist:1997gq}
  T.~Appelquist, A.~Nyffeler and S.~B.~Selipsky,
  ``Analyzing chiral symmetry breaking in supersymmetric gauge theories,''
  Phys.\ Lett.\  B {\bf 425}, 300 (1998)
  [arXiv:hep-th/9709177].


\bibitem{Sugimoto:1999tx}
  S.~Sugimoto,
  ``Anomaly cancellations in type I D9-D9-bar system and the USp(32)  string
  theory,''
  Prog.\ Theor.\ Phys.\  {\bf 102}, 685 (1999)
  [arXiv:hep-th/9905159].

\bibitem{Uranga:1999ib}
  A.~M.~Uranga,
  ``Comments on non-supersymmetric orientifolds at strong coupling,''
  JHEP {\bf 0002}, 041 (2000)
  [arXiv:hep-th/9912145].

\bibitem{Armoni:2007jt}
  A.~Armoni,
  ``Non-Perturbative Planar Equivalence and the Absence of Closed String
  Tachyons,''
  JHEP {\bf 0704}, 046 (2007)
  [arXiv:hep-th/0703229].

\end{thebibliography}
\end{document}